\documentclass[aps,prb,reprint,groupedaddress]{revtex4-1}

\usepackage{amsmath,amsthm,amssymb}
\usepackage{amsmath}
\usepackage{bm}
\usepackage[dvipdfm]{graphicx}
\usepackage{mathrsfs}

\begin{document}


\title{Spin Susceptibility in the Superconducting state of Ferromagnetic Superconductor UCoGe}


\author{T.~Hattori}
\altaffiliation{t.hattori@scphys.kyoto-u.ac.jp}
\author{K.~Karube}
\author{Y.~Ihara}
\altaffiliation{Present address: {\it Department of Physics, Graduate School of Science, Hokkaido University, Sapporo 060-0810, Japan}}
\author{K.~Ishida}
\affiliation{Department of Physics, Graduate School of Science, Kyoto University, Kyoto 606-8502, Japan}
\author{K.~Deguchi}
\author{N.~K.~Sato}
\affiliation{Department of Physics, Graduate School of Science, Nagoya University, Nagoya 464-8602, Japan}
\author{T.~Yamamura}
\affiliation{Institute for Materials Research, Tohoku University. Sendai 980-8577, Japan}


\newcommand{\UGe}{UGe$_2$}
\newcommand{\Co}{$^{59}$Co~}
\newcommand{\TCurie}{$T_{\rm Curie}~$}
\newcommand{\TSC}{$T_{\rm SC}~$}

\date{\today}

\begin{abstract}
In order to determine the superconducting paring state in the ferromagnetic superconductor UCoGe, $^{59}$Co NMR Knight-shift, which is directly related to the microscopic spin susceptibility, was measured in the superconducting state under magnetic fields perpendicular to spontaneous magnetization axis: $^{59}K^{a, b}$. $^{59}K^{a, b}$ shows to be constant, but does not decrease below a superconducting transition. These behaviors as well as the invariance of the internal field at the Co site in the superconducting state exclude the spin-singlet pairing, and can be interpreted with the equal-spin pairing state with a large exchange field along the $c$ axis, which was studied by Mineev [Phys. Rev. B 81, 180504 (2010)]. 
\end{abstract}


\maketitle


\section{INTRODUCTION \label{Introduction}}
 The discovery of superconductivity in itinerant ferromagnets has had a great impact on the community for studying superconductivity\cite{SSSaxena2000, DAoki2001, TAkazawa2004, NTHuy2007}, since they are considered as the most promising candidate of a spin-triplet superconductor. The intimate relationship between ferromagnetic (FM) fluctuations and superconductivity in UCoGe\cite{THattori2012PRL}, is a strong experimental suggestion of spin-triplet superconductivity. In an itinerant FM superconductor with the presence of a large energy splitting between the majority and minority spin Fermi surfaces, exotic spin-triplet superconductivity is anticipated from a theoretical point of view\cite{DFay1980}, in which pairing is between parallel spins within each spin Fermi surface. Since spin-triplet superconductivity possesses multiple internal degrees of freedom, the identification of the spin state is a first step in understanding the spin-triplet superconductors. 

Spin-susceptibility of the spin-triplet superconductor with the equal spin ($|\uparrow \uparrow \rangle$  or $|\downarrow \downarrow \rangle$) pairing along the spin-quantization axis keeps its normal state value if the spin-quantization axis follows the external-field direction. On the other hand, in the presence of the strong spin-orbit interaction fixing the mutual orientation of the spin-quantization axis and the crystalline symmetry direction, spin susceptibility shows anisotropic behavior: spin susceptibility parallel to the spin-quantization axis is unchanged, but spin susceptibility perpendicular to the axis vanishes  at $T = 0$. In uranium compounds, spin-orbit interaction is expected to be large since the magnetic anisotropy in the normal state is quite strong. To detect the variation of the spin susceptibility related to the superconducting (SC) pairing, Knight-shift measurement is one of a few experimental techniques to probe the spin susceptibility in the SC state, since Knight-shift measures the hyperfine field at the nuclear site produced by electrons.  

 Here we report on Knight-shift measurements in the SC state of a FM superconductor. We measured \Co Knight-shift ($^{59}K$) in the FM superconductor UCoGe.  Studies of the SC upper critical field ($H_{\rm c2}$) and its angle dependence along each crystalline axis revealed remarkable enigmatic behavior\cite{NTHuy2008, DAoki2009JPSJ} : superconductivity survives far beyond the Pauli-limiting field along the $a$ and $b$ axis, whereas $H_{\rm c2}$ along the magnetic easy axis ($c$ axis) is as small as 0.5 T. Since the NMR linewidth along the $c$ axis is so broad, and $H_{\rm c2}$ along the $c$ axis is so small that we could not detect the $^{59}$Co NMR signal for $H||c$ in the SC state, we focus on the $^{59}$Co Knight-shift measurements for $H || a$ and $b$.

\section{EXPERIMENT \label{Experiment}}
 The single crystalline UCoGe was utilized for the measurement, which is the same sample reported previously \cite{TOhta2010, THattori2012PRL}. The sample showed a large residual resistivity ratio ($RRR$) of approximately 30 along $b$ axis. The FM transition temperature was evaluated to be 2.55 K from the Arrot plots, and the midpoint SC transition temperatures were determined from $ac$ susceptibility as 0.57 K. Clear anomalies in the specific heat were observed at \TCurie and $T_{\rm SC}$, indicating that two anomalies are the bulk transitions. Microscopic measurements have shown the occurrence of superconductivity in the FM region of the sample, indicating homogeneous coexistence of ferromagnetism and superconductivity\cite{TOhta2010, ADevisser2009}. The NQR measurements on the present single-crystal sample indicates that nearly half region of the sample is in the SC state but remaining region is non-SC, although the whole region is in the FM state above $T_{\rm SC}$\cite{TOhta2010}. Here we stress that the non-SC region remaining below $T_{\rm SC}$ seems intrinsic in the FM superconductors since the clear relationship between the spontaneous moments and the non-SC fraction was observed\cite{DAokiJPSJS2011}.

\begin{figure}[t]
\includegraphics[width=\hsize,clip]{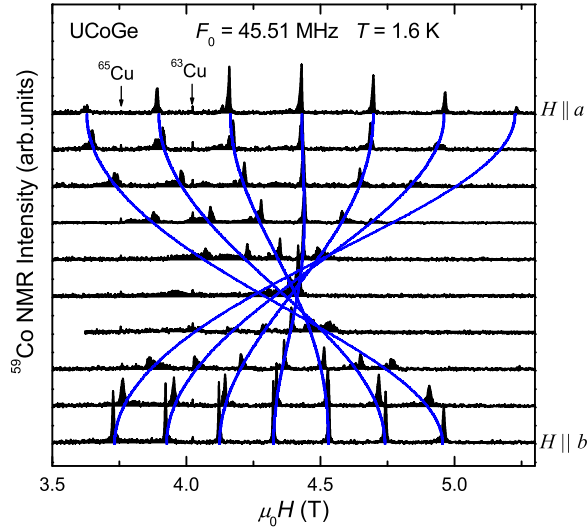}
\vspace*{-22pt}
\caption{(color online) Field-swept NMR spectra measured by using dilution refrigerator with fields rotated in $ab$ plane at $T = 1.6$ K, with the simulated locus of NMR peaks (7 line).}\label{Fig:Hsweep-ab-plane}
\end{figure}

The angle-dependent NMR measurements were performed using a split-coil SC magnet with a single-axis rotator. For the measurement at low temperatures, the $^3$He-$^4$He dilution refrigerator, in which the sample was mounted, was rotated against the split-coil SC magnet, and the single-crystal sample was immersed in  $^3$He-$^4$He mixture to avoid RF heating for the NMR measurements. The angle-dependent \Co NMR spectra obtained in the $ab$ plane with the dilution refrigerator are shown in the Fig. \ref{Fig:Hsweep-ab-plane}. Peak magnetic fields in each NMR spectrum are extracted by solving the secular equations of
\begin{align}
\mathcal{H} &= \mathcal{H}_Z + \mathcal{H}_Q\\
=&\gamma_n \hbar (1 + \bm{K}) \bm{I \cdot H} + \frac{\hbar \nu_{\text{Q}}}{6} \left\{  (3 I_z^2 -\bm{I}^2) + \frac{1}{2} \eta (I^2_+ + I^2_-)\right\}. \nonumber
\end{align}
Here $\mathcal{H}_Z$ and $\mathcal{H}_Q$ are the Zeeman and electric quadrupole Hamiltonian, $\bm{K}$ and $\bm{H}$ are the Knight-shift tensor and external-field vector, and $\nu_Q$ and $\eta$ are the electric quadrupole frequency and asymmetric parameter. Since the $\nu_Q$, $\eta$ and the direction of the principal axis of $\bm{I}$ to the crystal axis are determined already with the previous NQR/NMR measurements respectively\cite{TOhta2010, THattori2011JPSJS}, we can estimate the field direction to the crystal axis from the NMR peak locus. The satisfaction with simulation of peak locus in Fig.\ref{Fig:Hsweep-ab-plane} represents the fine tuning of field angle: the misalignment is estimated to be less than 2 degree. $^{59}K$ shown below were obtained from the central or satellite \Co peaks. 

The occurrence of superconductivity under zero and magnetic fields was monitored by the measurement of the Meissner signal with the NMR coil in the same condition of the $^{59}K$ measurement. Temperature dependence of the SC Meissner signal was measured with high frequency ($f_{\rm req}$) $ac$ magnetic susceptibility measurements by observing the tuning frequency of the NMR circuit near $20$ MHz. When the single-crystal UCoGe undergoes a SC transition, $\chi_{\rm bulk}$ becomes negative due to the Meissner effect and thus $f_{\rm req}$ increases in the SC state.

\section{Result and Discussion \label{Result}}
\begin{figure}[t]
\includegraphics[width=\hsize,clip]{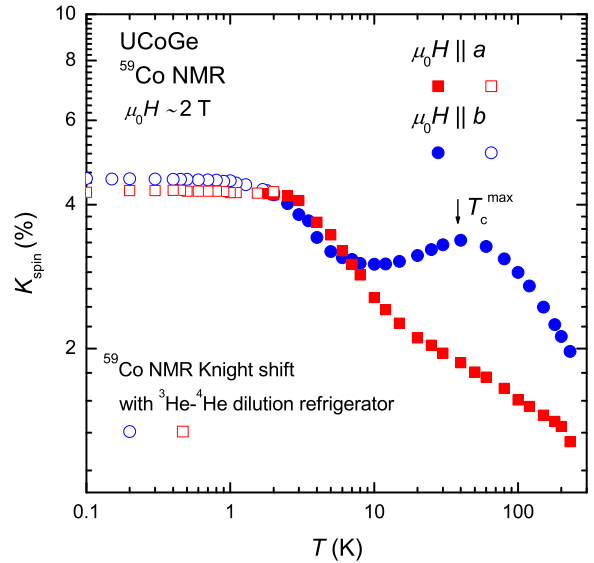}
\vspace*{-24pt}
\caption{(color online) Temperature dependence of Knight-shift $(K)$ measured in fields along the $a$ axis and $b$ axis under $\sim 2$ T. The outline shape under $T = 2$ K represents the result with using dilution refrigerator and the absolute value is slightly shifted to the $^{59}K$ in the high temperature region.}\label{Fig:Kab}
\end{figure}
Figure \ref{Fig:Kab} shows the temperature dependence of $^{59}K$ for $\mu_0 H || a, b \sim 2$ T. In the figure, we connect $^{59}K$ measured with the dilution refrigerator to $^{59}K$ in a high temperature region. $^{59}K$ for $H || a$ and $b$ in the high temperature region is nicely scaled to the bulk susceptibility $\chi$ measured for $H || a$ and $b$, as displayed in the previous paper\cite{YIhara2010PRL}. The slope of the relation is positive and is nearly the same in three directions, indicating that the \Co nuclear spins are largely affected by the U-$5f$ electronic spins through the hybridization between U-$5f$ and Co-$4s$ orbitals. $^{59}K$ for $H || b$ shows a broad maximum around 40 K. It was reported recently that similar peak was observed in $\chi$ for $H || b$ at $T_c^{\rm max} \sim 37.5$ K, and that the slope of magnetization $M(H)$ for $H || b$ shows a significant enhancement at $\mu_0 H_M^{\rm kink} = 46 \pm 2$ T\cite{WKnafo2012PRB}. Such a metamagnetic-like behavior has been reported in various heavy-fermion (HF) compounds. It is noteworthy that similar $T^{\rm max}$ and  $\mu_0 H_M^{\rm kink}$ for $H || b$ were reported to be 10 K and 12 T \cite{WKnafo2012PRB, FLevy2005URhGeReSC, AMiyake2008JPSJ} respectively in URhGe, and the ratio of $\mu_0 H_M^{\rm kink}/ T^{\rm max}$ is nearly the same between two compounds. In addition, since the resistivity along the $c$ axis turns to be metallic below around $T^{\rm max}$\cite{THattori2012PRL}, the metamagnetic energy scale is regarded as a characteristic energy of the HF state in UCoGe. 

\begin{figure}[t]
\includegraphics[width=\hsize,clip]{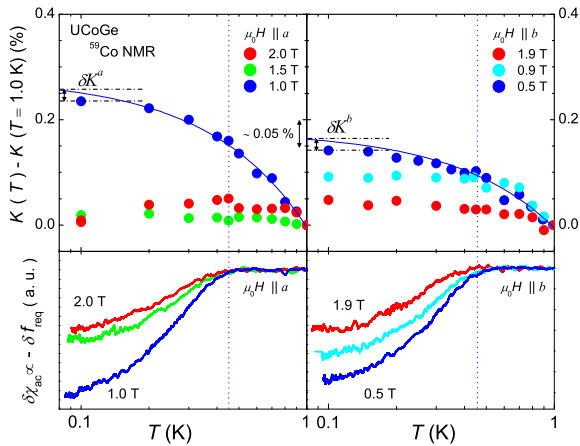}
\vspace*{-30pt}
\caption{(color online) Temperature dependence of \Co NMR Knight-shift and the Meissner signal for $H || a, b$ below 1 K. The change from the value at 1 K is represented. The blue line in the top two fig. is extrapolation of the linear fit between $T_{\rm SC}$ and 1 K.} \label{Fig:K-Meissner}
\vspace*{-5pt}
\end{figure}

Now, we move on the detail of $^{59}K$ in the SC state. Figure \ref{Fig:K-Meissner} shows the temperature dependence of $^{59}K$ and the Meissner signal below 1 K, measured under various fields. In Fig. 3, the deviation from $^{59}K$ at $T = 1$ K, $[\Delta K \equiv K - K(1~{\rm K})]$ is shown, since it was found from the NQR measurement\cite{TOhta2010} that the whole region of the same sample is in the FM state below 1 K. In the normal state, $\Delta K$ increases with decreasing temperature, following the development of the FM moments. At $\mu_0 H = 1$ T for $H || a$ and $\mu_0 H = 0.5$ T for $H || b$, the increase of  $^{59}K$ gets dull or saturates around the temperature below which the Meissner signal appears (vertical line), resulting in the derivation from the extrapolation of $K$ to $T = 0$ K as shown in Fig. \ref{Fig:K-Meissner}. The extrapolation of $K$ is determined from the linear fit on $K$ from 1 K to $T_{\rm SC}$, and would give the upper limit of $K$ at $T = 0$ K. Therefore the derivation from the extrapolation of $K$, $\delta K^{a,b}$ (less than 0.05 \%) is the maximum value of the suppression of $K$ due to the occurrence of superconductivity. 

The tiny presence or absence of the $^{59}K$ suppression below \TSC excludes the spin-singlet pairing, since appreciable decrease of the order of $10^{-1 \sim 0}$\% is expected in the spin-singlet pairing. However the tiny or absence of the $^{59}K$ suppression also seems to be incompatible with the equal spin-pairing state with the spin direction parallel to the $c$ axis, since spin susceptibility along the $a$ and $b$ axis should decrease below \TSC in this case. There is, however, a spontaneous magnetization ($M_c$) in the FM superconductor, which splits the up-spin and down-spin bands significantly. Recently, Mineev studied the equal spin-pairing state with a spin-quantization axis parallel to the direction of $M_c$, which is induced by the itinerant ferromagnet band splitting, and gave the microscopic derivation of the paramagnetic susceptibility in FM superconductors for the field perpendicular to $M_c$\cite{VPMineev2010PRB}. In his model, the normal-state susceptibility perpendicular to $M_c$, $\chi^n_{\perp}$ is expressed with the numbers of electrons in the spin-up and down band $N_{\uparrow, \downarrow}$ as 
\[
\chi_{\perp}^{n} = \mu_{\rm B} \frac{N_{\uparrow} - N_{\downarrow}}{h},
\]
where $h$ is the exchange field acting on the electron spins along $M_c$. Thus, the susceptibility related to the SC pairs originates from the electrons filling the momentum-space shell between the Fermi surfaces of the spin-up and spin-down bands, and the SC transition changes the Fermi distribution of the electrons only over energies close to the Fermi surfaces within an order $\Delta$. Thus, the suppression of the perpendicular component of the spin susceptibility at $T = 0$ due to the spin-triplet pairing is calculated as
\[
\delta \chi_{\perp} \equiv \chi_{\perp}^{s} (T=0) - \chi_{\perp}^n (0) = -a\chi_{\perp}^n \frac{\Delta^2}{(\mu_{\rm B} h)^2} \ln \frac{\mu_{\rm B} h}{|\Delta|},
\]
where $|\Delta|$ is the characteristic quantity of the gap amplitudes and $a$ is a numerical constant\cite{VPMineev2010PRB}. To estimate the suppression of the spin susceptibility, we need to know the value of the exchange field along the $c$ axis, $h$. Quite recently, the magnetization $M(H)$ of UCoGe was measured at 1.5 K up to 60 T, and the data indicates that $M(H)$ along the $c$ axis is roughly denoted as
\[
M_c (H) \sim \left( \frac{\partial M_c}{\partial H} \right) H + M_c (0),
\]
with ($\frac{\partial M_c}{\partial H}$) nearly constant (0.02 $\mu_{\rm B}$T$^{-1}$) in the $H$ range from 5 to 15 T\cite{WKnafo2012PRB}. If we assume this relation and use the magnetization value at zero external field $M_c \sim 0.07 \mu_{\rm B}$, the exchange field along the $c$ axis, $h$ is estimated to be 0.07($\mu_{\rm B})/0.02(\mu_{\rm B}$ T$^{-1}$)$ = 3.5$ T. It should be noted that the estimated $h$ is a minimum field, since $h$ can become larger by the electron correlation effect. Adopting the estimated $h, |\Delta|/k_{\rm B} \sim$ \TSC $=0.6$ K and $a \sim 1$, the suppression $\delta \chi_{\perp} / \chi^n_{\perp} $ is estimated to be $\le 0.06$, which should be compared with the experimental results.

 As reported in the previous paper, the NQR measurement on the present single-crystal sample indicates nearly half fraction in the non-SC state. Thus the suppression ratio of the $^{59}K$ ascribed to superconductivity is roughly estimated as 
$ \delta K^{a,b}/K^{a,b} \sim 0.05 \% / (4 \% \times 0.5) = 0.025$, which is in rough agreement with the crude estimation based on the spin-triplet pairing. It is, however, difficult to insist that there is actually a small decrease or kink  because this estimated value is so small, almost comparable with experimental error of the order $10^{-2}$ \%.

\begin{figure}[tbp]
\includegraphics[width=\hsize,clip]{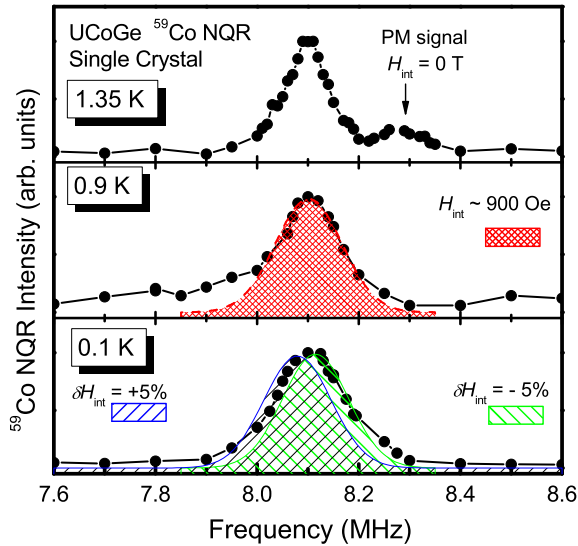}
\vspace*{-20pt}
\caption{(color online)$^{59}$Co NQR spectra from the $\pm 5/2 \leftrightarrow \pm 7/2$ transitions measured at 1.35 K and 0.9 K ($< T_{\rm Curie}$), and 0.1 K ($< T_{\rm SC}$)\cite{TOhta2010}. In the NQR spectrum at 1.35 K, small paramagnetic (PM) state signal is observed with the main FM-state signal. In the NQR spectrum at 0.9 K, the PM-state signal is not observed and the calculated NQR spectrum with $H_{\rm int}\sim 900$ Oe and the Gaussian distribution is shown by the red hatch. In the NQR spectrum at 0.1 K, the calculated NQR spectra with $\delta H_{\rm int}$ is $\pm5$ \% $H_{\rm int}$ are shown for the estimation of  possible change of $H_{\rm int}$.}\label{Fig:Fig-Hint-TDep}
\vspace*{-5pt}
\end{figure}

In the present measurements, we could not measure $^{59}K$ for $H || c$ in the SC state as mentioned above. We reported, however, from the NQR measurements that the internal field at the Co site $H_{\rm int}$ is unchanged passing through \TSC (Fig.\ref{Fig:Fig-Hint-TDep}$^[$\cite{TOhta2010}$^]$), indicative of the invariance of the spontaneous moment in the SC state, since $H_{\rm int}$ arises from the spontaneous ordered moment $M_c$ with the relation of $H_{\rm int} \propto M_c$. From the simulation of the NQR spectrum at 0.1 K shown in Fig.\ref{Fig:Fig-Hint-TDep}, the change of the internal field below \TSC normalized with the internal field above $T_{\rm SC}$ $\delta H_{\rm int}/H_{\rm int}^{\rm n}  = |H_{\rm int}^{\rm n}-H_{\rm int}^{\rm s}|/H_{\rm int}^{\rm n}$ is less than $\pm 5 ~\%$ if any. The absent of large decrease of internal field is also reported by $\mu$SR experiment\cite{ADevisser2009}. Mineev calculated $M_c$ in the SC state ($M_c^s$), and the difference from its normal state value ($\delta M_c = M_c^s - M_c^n$) is expressed as
\[
 \delta M_c = M_c^n \frac{\left(D_{\uparrow}' |\Delta_{\uparrow}|^2 -D_{\downarrow}' |\Delta_{\downarrow}|^2 \right)}{N_{\uparrow} - N_{\downarrow}} \ln \frac{\epsilon_F }{T_{\rm SC}}.
\]
$D_{\uparrow {\rm or} \downarrow}$ is the density of states at the Fermi level for the spin-up and down band, respectively, and  
$D_{\uparrow {\rm or} \downarrow}'$ is the energy derivatives of them \cite{VPMineev2010PRB}. If we assume $|\Delta_{\uparrow}| \sim |\Delta_{\downarrow}|$ and $D_{\uparrow \downarrow}' \sim D_{\uparrow \downarrow}/\epsilon_F \sim N_{\uparrow \downarrow}/\epsilon_F^2$  for order estimation,
$\delta M_c^{\rm s} / M_c^{\rm n} \sim |\Delta|^2/\epsilon_F^2 \ln (\epsilon_F / T_{\rm SC})\sim 10^{-3}$ is evaluated. 
Here we take $\epsilon_F/k_{\rm B} \sim 40$ K, which is a characteristic temperature of U-$5f$ moments. Although the very tiny decrease in $H_{\rm int}$ estimated by Mineev's theory is undetectable with the resolution of the present NQR experiment, the absence of appreciable change in the SC state can be interpreted by the triplet-pairing scenario consistently.

Finally, we point out the similarity between FM superconductors with the Ising anisotropy along the $c$ axis and an inversion-symmetry breaking superconductor along the $c$ axis. In the latter superconductor, e.g. CeIrSi$_3$, two Fermi surfaces are split with momentum $\bm{k}$ dependence due to the strong antisymmetric spin-orbit coupling effect originating from the Rashba-type interaction, which is proportional to an antisymmetric potential gradient $\nabla V$ ($\parallel c$). 
In this case, spins on each Fermi surface are parallel to $\mbox{\boldmath$k$} \times \nabla V$, which lie in the $ab$ plane. 
The spin susceptibility in the SC state $\chi_s^{\rm SC}$ was theoretically shown to be unchanged in $\mbox{\boldmath$H$} \parallel c$, but to decrease in $\mbox{\boldmath$H$} \perp c$ with dependency on the strength of the spin-orbit interaction, when the Rashba-type interaction is sufficiently larger than the SC gap.
 The spin-susceptibility behavior in the SC state on CeIrSi$_3$ was actually measured with the $^{27}$Si NMR Knight-shift\cite{HMukuda2010PRL}, and is in good agreement with the theoretical prediction. We point out that the almost constant Knight-shift on UCoGe in $\mbox{\boldmath$H$} \perp c$ is similar to that on the noncentrosymmetric superconductor in $\mbox{\boldmath$H$} \parallel c$, since the Fermi-surface splitting is larger than the SC gap and $\mbox{\boldmath$H$}$ is applied perpendicular to the spin component in both superconductors. It is noteworthy that the field direction perpendicular to the spin component largely exceeds the Pauli limiting field in two superconductors. This also suggests that the change of spin susceptibility below \TSC is small or absent.

\section{Summary \label{summary}}
  In conclusion, we measured the \Co Knight-shift for $H || a$ and $b$ in the SC state on the FM superconductor UCoGe, and found the almost constant behavior below \TSC in the \Co Knight-shift. The observed Knight-shift results as well as unchanged spontaneous moments in the SC state exclude the spin-singlet pairing, and can be reasonably interpreted with the spin triplet with a band splitting scenario where equal spin pairs with a spin quantization axis parallel to the direction of spontaneous magnetization and the band splitting energy is larger than the SC gap energy.

\begin{acknowledgments}
The authors thank S. Yonezawa, and Y. Maeno for experimental support and valuable discussions, and Y. Tada, S. Fujimoto, A. de Visser, D. Aoki, and J. Flouquet for valuable discussions. This work was partially supported by Kyoto Univ. LTM center, Yukawa Institute, the "Heavy Electrons" Grant-in-Aid for Scientific Research on Innovative Areas  (No. 20102006 and No. 23244075) from The Ministry of Education, Culture, Sports, Science, and Technology (MEXT) of Japan, a Grant-in-Aid for the Global COE Program ``The Next Generation of Physics, Spun from Universality and Emergence'' from MEXT of Japan, a grant-in-aid for Scientific Research from Japan Society for Promotion of Science (JSPS), KAKENHI (S) (No. 20224015).
\end{acknowledgments}

\bibliographystyle{apsrev}
\bibliography{Ref.bib}

\begin{thebibliography}{18}
\expandafter\ifx\csname natexlab\endcsname\relax\def\natexlab#1{#1}\fi
\expandafter\ifx\csname bibnamefont\endcsname\relax
  \def\bibnamefont#1{#1}\fi
\expandafter\ifx\csname bibfnamefont\endcsname\relax
  \def\bibfnamefont#1{#1}\fi
\expandafter\ifx\csname citenamefont\endcsname\relax
  \def\citenamefont#1{#1}\fi
\expandafter\ifx\csname url\endcsname\relax
  \def\url#1{\texttt{#1}}\fi
\expandafter\ifx\csname urlprefix\endcsname\relax\def\urlprefix{URL }\fi
\providecommand{\bibinfo}[2]{#2}
\providecommand{\eprint}[2][]{\url{#2}}

\bibitem[{\citenamefont{Saxena et~al.}(2000)\citenamefont{Saxena, Agarwal,
  Ahilan, Grosche, Haselwimmer, Steiner, Pugh, Walker, Julian, Monthoux
  et~al.}}]{SSSaxena2000}
\bibinfo{author}{\bibfnamefont{S.~S.} \bibnamefont{Saxena}},
  \bibinfo{author}{\bibfnamefont{P.}~\bibnamefont{Agarwal}},
  \bibinfo{author}{\bibfnamefont{K.}~\bibnamefont{Ahilan}},
  \bibinfo{author}{\bibfnamefont{F.~M.} \bibnamefont{Grosche}},
  \bibinfo{author}{\bibfnamefont{R.~K.~W.} \bibnamefont{Haselwimmer}},
  \bibinfo{author}{\bibfnamefont{M.~J.} \bibnamefont{Steiner}},
  \bibinfo{author}{\bibfnamefont{E.}~\bibnamefont{Pugh}},
  \bibinfo{author}{\bibfnamefont{I.~R.} \bibnamefont{Walker}},
  \bibinfo{author}{\bibfnamefont{S.~R.} \bibnamefont{Julian}},
  \bibinfo{author}{\bibfnamefont{P.}~\bibnamefont{Monthoux}},
  \bibnamefont{et~al.}, \bibinfo{journal}{Nature}
  \textbf{\bibinfo{volume}{406}}, \bibinfo{pages}{587} (\bibinfo{year}{2000}).

\bibitem[{\citenamefont{Aoki et~al.}(2001)\citenamefont{Aoki, A.Huxley,
  E.Ressouche, D.Braithwaite, J.Flouquet, Brison, E.Lhotel, and
  C.Paulsen}}]{DAoki2001}
\bibinfo{author}{\bibfnamefont{D.}~\bibnamefont{Aoki}},
  \bibinfo{author}{\bibnamefont{A.Huxley}},
  \bibinfo{author}{\bibnamefont{E.Ressouche}},
  \bibinfo{author}{\bibnamefont{D.Braithwaite}},
  \bibinfo{author}{\bibnamefont{J.Flouquet}},
  \bibinfo{author}{\bibfnamefont{J.}~\bibnamefont{Brison}},
  \bibinfo{author}{\bibnamefont{E.Lhotel}}, \bibnamefont{and}
  \bibinfo{author}{\bibnamefont{C.Paulsen}}, \bibinfo{journal}{Nature}
  \textbf{\bibinfo{volume}{413}}, \bibinfo{pages}{613} (\bibinfo{year}{2001}).

\bibitem[{\citenamefont{Akazawa et~al.}(2004)\citenamefont{Akazawa, Hidaka,
  Fujiwara, Kobayashi, Yamamoto, Haga, Settai, and
  $\bar{\text{O}}$nuki}}]{TAkazawa2004}
\bibinfo{author}{\bibfnamefont{T.}~\bibnamefont{Akazawa}},
  \bibinfo{author}{\bibfnamefont{H.}~\bibnamefont{Hidaka}},
  \bibinfo{author}{\bibfnamefont{T.}~\bibnamefont{Fujiwara}},
  \bibinfo{author}{\bibfnamefont{T.~C.} \bibnamefont{Kobayashi}},
  \bibinfo{author}{\bibfnamefont{E.}~\bibnamefont{Yamamoto}},
  \bibinfo{author}{\bibfnamefont{Y.}~\bibnamefont{Haga}},
  \bibinfo{author}{\bibfnamefont{R.}~\bibnamefont{Settai}}, \bibnamefont{and}
  \bibinfo{author}{\bibfnamefont{Y.}~\bibnamefont{$\bar{\text{O}}$nuki}},
  \bibinfo{journal}{J. Phys.: Condens. Matter} \textbf{\bibinfo{volume}{16}},
  \bibinfo{pages}{L29} (\bibinfo{year}{2004}).

\bibitem[{\citenamefont{Huy et~al.}(2007)\citenamefont{Huy, Gasparini, de~Nijs,
  Huang, Klaasse, Gortenmulder, de~Visser, Hamann, G{\"{o}}rlach, and
  v.~L{\"{o}}hneysen}}]{NTHuy2007}
\bibinfo{author}{\bibfnamefont{N.~T.} \bibnamefont{Huy}},
  \bibinfo{author}{\bibfnamefont{A.}~\bibnamefont{Gasparini}},
  \bibinfo{author}{\bibfnamefont{D.~E.} \bibnamefont{de~Nijs}},
  \bibinfo{author}{\bibfnamefont{Y.}~\bibnamefont{Huang}},
  \bibinfo{author}{\bibfnamefont{J.~C.~P.} \bibnamefont{Klaasse}},
  \bibinfo{author}{\bibfnamefont{T.}~\bibnamefont{Gortenmulder}},
  \bibinfo{author}{\bibfnamefont{A.}~\bibnamefont{de~Visser}},
  \bibinfo{author}{\bibfnamefont{A.}~\bibnamefont{Hamann}},
  \bibinfo{author}{\bibfnamefont{T.}~\bibnamefont{G{\"{o}}rlach}},
  \bibnamefont{and}
  \bibinfo{author}{\bibfnamefont{H.}~\bibnamefont{v.~L{\"{o}}hneysen}},
  \bibinfo{journal}{Phys. Rev. Lett.} \textbf{\bibinfo{volume}{99}},
  \bibinfo{pages}{067006} (\bibinfo{year}{2007}).

\bibitem[{\citenamefont{Hattori et~al.}(2012)\citenamefont{Hattori, Ihara,
  Nakai, Ishida, Tada, Fujimoto, Kawakami, Osaki, Deguchi, Sato
  et~al.}}]{THattori2012PRL}
\bibinfo{author}{\bibfnamefont{T.}~\bibnamefont{Hattori}},
  \bibinfo{author}{\bibfnamefont{Y.}~\bibnamefont{Ihara}},
  \bibinfo{author}{\bibfnamefont{Y.}~\bibnamefont{Nakai}},
  \bibinfo{author}{\bibfnamefont{K.}~\bibnamefont{Ishida}},
  \bibinfo{author}{\bibfnamefont{Y.}~\bibnamefont{Tada}},
  \bibinfo{author}{\bibfnamefont{S.}~\bibnamefont{Fujimoto}},
  \bibinfo{author}{\bibfnamefont{N.}~\bibnamefont{Kawakami}},
  \bibinfo{author}{\bibfnamefont{E.}~\bibnamefont{Osaki}},
  \bibinfo{author}{\bibfnamefont{K.}~\bibnamefont{Deguchi}},
  \bibinfo{author}{\bibfnamefont{N.~K.} \bibnamefont{Sato}},
  \bibnamefont{et~al.}, \bibinfo{journal}{Phys. Rev. Lett.}
  \textbf{\bibinfo{volume}{108}}, \bibinfo{pages}{066403}
  (\bibinfo{year}{2012}).

\bibitem[{\citenamefont{Fay and Appel}(1980)}]{DFay1980}
\bibinfo{author}{\bibfnamefont{D.}~\bibnamefont{Fay}} \bibnamefont{and}
  \bibinfo{author}{\bibfnamefont{J.}~\bibnamefont{Appel}},
  \bibinfo{journal}{Phys. Rev. B} \textbf{\bibinfo{volume}{22}},
  \bibinfo{pages}{3173} (\bibinfo{year}{1980}).

\bibitem[{\citenamefont{Huy et~al.}(2008)\citenamefont{Huy, de~Nijs, Huang, and
  de~Visser}}]{NTHuy2008}
\bibinfo{author}{\bibfnamefont{N.~T.} \bibnamefont{Huy}},
  \bibinfo{author}{\bibfnamefont{D.~E.} \bibnamefont{de~Nijs}},
  \bibinfo{author}{\bibfnamefont{Y.~K.} \bibnamefont{Huang}}, \bibnamefont{and}
  \bibinfo{author}{\bibfnamefont{A.}~\bibnamefont{de~Visser}},
  \bibinfo{journal}{Phys. Rev. Lett.} \textbf{\bibinfo{volume}{100}},
  \bibinfo{pages}{077002} (\bibinfo{year}{2008}).

\bibitem[{\citenamefont{Aoki et~al.}(2009)\citenamefont{Aoki, Matsuda, Taufour,
  Hassinger, Knebel, and Flouquet}}]{DAoki2009JPSJ}
\bibinfo{author}{\bibfnamefont{D.}~\bibnamefont{Aoki}},
  \bibinfo{author}{\bibfnamefont{T.~D.} \bibnamefont{Matsuda}},
  \bibinfo{author}{\bibfnamefont{V.}~\bibnamefont{Taufour}},
  \bibinfo{author}{\bibfnamefont{E.}~\bibnamefont{Hassinger}},
  \bibinfo{author}{\bibfnamefont{G.}~\bibnamefont{Knebel}}, \bibnamefont{and}
  \bibinfo{author}{\bibfnamefont{J.}~\bibnamefont{Flouquet}},
  \bibinfo{journal}{J. Phys. Soc. Jpn.} \textbf{\bibinfo{volume}{78}},
  \bibinfo{pages}{113709} (\bibinfo{year}{2009}).

\bibitem[{\citenamefont{Ohta et~al.}(2010)\citenamefont{Ohta, Hattori, Ishida,
  Nakai, Osaki, Deguchi, Sato, and Satoh}}]{TOhta2010}
\bibinfo{author}{\bibfnamefont{T.}~\bibnamefont{Ohta}},
  \bibinfo{author}{\bibfnamefont{T.}~\bibnamefont{Hattori}},
  \bibinfo{author}{\bibfnamefont{K.}~\bibnamefont{Ishida}},
  \bibinfo{author}{\bibfnamefont{Y.}~\bibnamefont{Nakai}},
  \bibinfo{author}{\bibfnamefont{E.}~\bibnamefont{Osaki}},
  \bibinfo{author}{\bibfnamefont{K.}~\bibnamefont{Deguchi}},
  \bibinfo{author}{\bibfnamefont{N.~K.} \bibnamefont{Sato}}, \bibnamefont{and}
  \bibinfo{author}{\bibfnamefont{I.}~\bibnamefont{Satoh}}, \bibinfo{journal}{J.
  Phys. Soc. Jpn.} \textbf{\bibinfo{volume}{79}}, \bibinfo{pages}{023707}
  (\bibinfo{year}{2010}).

\bibitem[{\citenamefont{de~Visser et~al.}(2009)\citenamefont{de~Visser, Huy,
  Gasparini, de~Nijs, Andreica, Baines, and Amato}}]{ADevisser2009}
\bibinfo{author}{\bibfnamefont{A.}~\bibnamefont{de~Visser}},
  \bibinfo{author}{\bibfnamefont{N.~T.} \bibnamefont{Huy}},
  \bibinfo{author}{\bibfnamefont{A.}~\bibnamefont{Gasparini}},
  \bibinfo{author}{\bibfnamefont{D.~E.} \bibnamefont{de~Nijs}},
  \bibinfo{author}{\bibfnamefont{D.}~\bibnamefont{Andreica}},
  \bibinfo{author}{\bibfnamefont{C.}~\bibnamefont{Baines}}, \bibnamefont{and}
  \bibinfo{author}{\bibfnamefont{A.}~\bibnamefont{Amato}},
  \bibinfo{journal}{Phys. Rev. Lett.} \textbf{\bibinfo{volume}{102}},
  \bibinfo{pages}{167003} (\bibinfo{year}{2009}).

\bibitem[{\citenamefont{Aoki et~al.}(2011)\citenamefont{Aoki, Matsuda, Hardy,
  Meingast, Taufour, Hassinger, Sheikin, Paulsen, Knebel, Kotegawa
  et~al.}}]{DAokiJPSJS2011}
\bibinfo{author}{\bibfnamefont{D.}~\bibnamefont{Aoki}},
  \bibinfo{author}{\bibfnamefont{T.~D.} \bibnamefont{Matsuda}},
  \bibinfo{author}{\bibfnamefont{F.}~\bibnamefont{Hardy}},
  \bibinfo{author}{\bibfnamefont{C.}~\bibnamefont{Meingast}},
  \bibinfo{author}{\bibfnamefont{V.}~\bibnamefont{Taufour}},
  \bibinfo{author}{\bibfnamefont{E.}~\bibnamefont{Hassinger}},
  \bibinfo{author}{\bibfnamefont{I.}~\bibnamefont{Sheikin}},
  \bibinfo{author}{\bibfnamefont{C.}~\bibnamefont{Paulsen}},
  \bibinfo{author}{\bibfnamefont{G.}~\bibnamefont{Knebel}},
  \bibinfo{author}{\bibfnamefont{H.}~\bibnamefont{Kotegawa}},
  \bibnamefont{et~al.}, \bibinfo{journal}{J. Phys. Spc. Jpn.}
  \textbf{\bibinfo{volume}{80}}, \bibinfo{pages}{SA008} (\bibinfo{year}{2011}).

\bibitem[{\citenamefont{Hattori et~al.}(2011)\citenamefont{Hattori, Ihara,
  Ishida, Nakai, Osaki, Deguchi, Sato, and Satoh}}]{THattori2011JPSJS}
\bibinfo{author}{\bibfnamefont{T.}~\bibnamefont{Hattori}},
  \bibinfo{author}{\bibfnamefont{Y.}~\bibnamefont{Ihara}},
  \bibinfo{author}{\bibfnamefont{K.}~\bibnamefont{Ishida}},
  \bibinfo{author}{\bibfnamefont{Y.}~\bibnamefont{Nakai}},
  \bibinfo{author}{\bibfnamefont{E.}~\bibnamefont{Osaki}},
  \bibinfo{author}{\bibfnamefont{K.}~\bibnamefont{Deguchi}},
  \bibinfo{author}{\bibfnamefont{N.~K.} \bibnamefont{Sato}}, \bibnamefont{and}
  \bibinfo{author}{\bibfnamefont{I.}~\bibnamefont{Satoh}}, \bibinfo{journal}{J.
  Phys. Soc. Jpn. Suppl.} \textbf{\bibinfo{volume}{80}}, \bibinfo{pages}{SA077}
  (\bibinfo{year}{2011}).

\bibitem[{\citenamefont{Ihara et~al.}(2010)\citenamefont{Ihara, Hattori,
  Ishida, Nakai, Osaki, Deguchi, Sato, and Satoh}}]{YIhara2010PRL}
\bibinfo{author}{\bibfnamefont{Y.}~\bibnamefont{Ihara}},
  \bibinfo{author}{\bibfnamefont{T.}~\bibnamefont{Hattori}},
  \bibinfo{author}{\bibfnamefont{K.}~\bibnamefont{Ishida}},
  \bibinfo{author}{\bibfnamefont{Y.}~\bibnamefont{Nakai}},
  \bibinfo{author}{\bibfnamefont{E.}~\bibnamefont{Osaki}},
  \bibinfo{author}{\bibfnamefont{K.}~\bibnamefont{Deguchi}},
  \bibinfo{author}{\bibfnamefont{N.~K.} \bibnamefont{Sato}}, \bibnamefont{and}
  \bibinfo{author}{\bibfnamefont{I.}~\bibnamefont{Satoh}},
  \bibinfo{journal}{Phys. Rev. Lett.} \textbf{\bibinfo{volume}{105}},
  \bibinfo{pages}{206403} (\bibinfo{year}{2010}).

\bibitem[{\citenamefont{Knafo et~al.}(2012)\citenamefont{Knafo, Matsuda, Aoki,
  Hardy, Scheerer, Ballon, Nardone, Zitouni, Meingast, and
  Flouquet}}]{WKnafo2012PRB}
\bibinfo{author}{\bibfnamefont{W.}~\bibnamefont{Knafo}},
  \bibinfo{author}{\bibfnamefont{T.~D.} \bibnamefont{Matsuda}},
  \bibinfo{author}{\bibfnamefont{D.}~\bibnamefont{Aoki}},
  \bibinfo{author}{\bibfnamefont{F.}~\bibnamefont{Hardy}},
  \bibinfo{author}{\bibfnamefont{G.~W.} \bibnamefont{Scheerer}},
  \bibinfo{author}{\bibfnamefont{G.}~\bibnamefont{Ballon}},
  \bibinfo{author}{\bibfnamefont{M.}~\bibnamefont{Nardone}},
  \bibinfo{author}{\bibfnamefont{A.}~\bibnamefont{Zitouni}},
  \bibinfo{author}{\bibfnamefont{C.}~\bibnamefont{Meingast}}, \bibnamefont{and}
  \bibinfo{author}{\bibfnamefont{J.}~\bibnamefont{Flouquet}},
  \bibinfo{journal}{Phys. Rev. B} \textbf{\bibinfo{volume}{86}},
  \bibinfo{pages}{184416} (\bibinfo{year}{2012}).

\bibitem[{\citenamefont{Levy et~al.}(2005)\citenamefont{Levy, Sheikin, Grenier,
  and Huxley}}]{FLevy2005URhGeReSC}
\bibinfo{author}{\bibfnamefont{F.}~\bibnamefont{Levy}},
  \bibinfo{author}{\bibfnamefont{I.}~\bibnamefont{Sheikin}},
  \bibinfo{author}{\bibfnamefont{B.}~\bibnamefont{Grenier}}, \bibnamefont{and}
  \bibinfo{author}{\bibfnamefont{A.~D.} \bibnamefont{Huxley}},
  \bibinfo{journal}{Science} \textbf{\bibinfo{volume}{309}},
  \bibinfo{pages}{1343} (\bibinfo{year}{2005}).

\bibitem[{\citenamefont{A.Miyake et~al.}(2008)\citenamefont{A.Miyake, Aoki, and
  Flouquet}}]{AMiyake2008JPSJ}
\bibinfo{author}{\bibnamefont{A.Miyake}},
  \bibinfo{author}{\bibfnamefont{D.}~\bibnamefont{Aoki}}, \bibnamefont{and}
  \bibinfo{author}{\bibfnamefont{J.}~\bibnamefont{Flouquet}},
  \bibinfo{journal}{Phys. Soc. Jpn.} \textbf{\bibinfo{volume}{77}},
  \bibinfo{pages}{094709} (\bibinfo{year}{2008}).

\bibitem[{\citenamefont{Mineev}(2010)}]{VPMineev2010PRB}
\bibinfo{author}{\bibfnamefont{V.~P.} \bibnamefont{Mineev}},
  \bibinfo{journal}{Phys. Rev. B} \textbf{\bibinfo{volume}{81}},
  \bibinfo{pages}{180504} (\bibinfo{year}{2010}).

\bibitem[{\citenamefont{Mukuda et~al.}(2010)\citenamefont{Mukuda, Ohara,
  Yashima, Kitaoka, Settai, Onuki, Itoh, and Haller}}]{HMukuda2010PRL}
\bibinfo{author}{\bibfnamefont{H.}~\bibnamefont{Mukuda}},
  \bibinfo{author}{\bibfnamefont{T.}~\bibnamefont{Ohara}},
  \bibinfo{author}{\bibfnamefont{M.}~\bibnamefont{Yashima}},
  \bibinfo{author}{\bibfnamefont{Y.}~\bibnamefont{Kitaoka}},
  \bibinfo{author}{\bibfnamefont{R.}~\bibnamefont{Settai}},
  \bibinfo{author}{\bibfnamefont{Y.}~\bibnamefont{Onuki}},
  \bibinfo{author}{\bibfnamefont{K.~M.} \bibnamefont{Itoh}}, \bibnamefont{and}
  \bibinfo{author}{\bibfnamefont{E.~E.} \bibnamefont{Haller}},
  \bibinfo{journal}{Phys. Rev. Lett.} \textbf{\bibinfo{volume}{104}},
  \bibinfo{pages}{017002} (\bibinfo{year}{2010}).

\end{thebibliography}

\end{document}